\documentclass[a4paper,12pt]{article}
\usepackage[utf8]{inputenc}

\usepackage{authblk}
 \usepackage{amssymb}
\usepackage{hyperref}
\usepackage{amsmath,amsfonts}
\usepackage[dvips]{graphics}
\usepackage[dvips]{graphicx}
\usepackage{subfigure}
\usepackage{epsfig}
\usepackage{epstopdf}
\usepackage{amsmath}
\usepackage{amsfonts}
\RequirePackage{color}
 \DeclareUnicodeCharacter{2212}{-}
\begin{document}
\title{Galactic Clustering Under   Power-law Modified Newtonian Potential}
\author[1]{Abdul W. Khanday\thanks{abdulwakeelkhanday@gmail.com} }

   \author[2,3,4,5]{Sudhaker Upadhyay\thanks{sudhakerupadhyay@gmail.com; sudhaker@associates.iucaa.in}}

  \author[1]{Prince A. Ganai\thanks{princeganai@nitsri.com} }
  
\affil[1]{{Department of Physics, National Institute of Technology,
Srinagar, Kashmir-190006, India} }

\affil[2]{Department of Physics, K.L.S. College, Nawada, 
Bihar 805110, India}
 \affil[3]{Department of Physics, Magadh University, Bodh Gaya,
 Bihar  824234, India}
\affil[4]{Inter-University Centre for Astronomy and Astrophysics (IUCAA) Pune, Maharashtra 411007, India}
\affil[5]{School of Physics, Damghan University, P. O. Box 3671641167, Damghan, Iran}
\date{}
  \maketitle		 
\begin{abstract}
We estimate galaxy clustering under a modified gravitational potential. In particular, 
the modifications in gravitational potential energy occur due to a power-law and cosmological
constant terms. We derive a canonical partition function for the   system of galaxies interacting under such a modified gravitational  potential. Moreover, we compute various thermodynamical equation of states for the system. We do comparative analysis in order to  emphasize the effect of corrections on thermodynamics of the system. Interestingly, the modifications in thermodynamical quantities are embedded in clustering parameter only.
\end{abstract}

	\section{Introduction}
	Since  last decade, the substantial progresses have been made in the understandings of galaxy clusters, from their internal structure and evolution to their place in the large scale structure of the universe. All these progresses are due to the stupendous improvements in the theoretical modeling and numerical simulation, viz-a-viz abundance of new information provided by multi-wavelength surveys of the universe. Various theories of cosmological many-body distribution function have been developed from the thermodynamic point of view.
 	Benkestien~\cite{1}, Hawking~\cite{2} and Unruh~\cite{3} originated the relation between relativity and thermodynamics. Later, Jacobson~\cite{4} introduced the Einstein equation as thermodynamic equation of state. In fact, the Einstein equation is derived  from the proportionality of entropy and the horizon area together with the fundamental relation $\delta Q = T dS$, where $\delta Q$ and $T$ are interpreted as the energy flux and Unruh temperature. Recently Verlinde~\cite{5} proposed an entropic origin of gravity and interpreted gravity as an entropic force. In this regard, it is argued that the central notion needed to derive gravity is information. Employing the holographic principle and equipartition law of energy,  Newton's law of gravitation, Poisson's equation and Einstein's field equations  are successfully obtained. Verlinde had used area-entropy relation of black holes in Einstein's gravity (i.e. $S = \frac{A}{4l_p^2}$) to  get the Newtons law, where $S, A$ and  $l_P$  refer the entropy of the black hole,  the area of horizon and  the Planck length, respectively.

The study of modified gravity on the cosmological scales is an active area of present research. Recently,  the corrected gravitational potential
plays a vital role in estimation of the total mass of a sample of 12 clusters of galaxies
 which  provides a better fit to the mass of visible matter   
  \cite{cap}. At large distances, the modification in  Newtonian potential occurs due
to the propagation of gravity into the bulk \cite{flo}.	
	Recently,  Sheykhi and  Hendi studied the effect of power-law corrections in entropy to the Newton's law \cite{12}. These corrections appear due to entanglement of quantum field inside and outside of horizon~\cite{6}. It is also found that a viable source of black hole entropy is quantum entanglement of degrees of freedom inside and outside the horizon. Also, the black hole entropy is found directly proportional to the surface area of the sphere when the field is in the ground state.  But when the field is in a superposition of ground and excited states, a correction term proportional to a fractional power of area appears. These corrections are negligible for large  horizon areas. Another quantum correction to Newton's law, a logarithmic correction, has also been studied which  appears due to the result of thermal equilibrium fluctuations and the quantum fluctuations of loop quantum gravity~\cite{7,8,9}. Recently, the effect of this logarithmic correction on the galaxy clustering has also been studied ~\cite{10}. The clustering of galaxies has been studied under
the modified gravity under various potentials  \cite{ san, mils, sud, MNRAS-Pour,PRD-Sud, MNRAS-Sud, sud1, Dark-Sud}.

 The power-law corrected entropy leads to modification in Newton's  gravitational potential by adopting the viewpoint of gravity as an entropic force. 
 We consider a power-law corrected entropy of the following form ~\cite{6,11}:
	\begin{equation}
	S = \dfrac{A}{4 l_p^2}[1-K_\alpha A^{1-\frac{\alpha}{2}}],\label{en}
 	\end{equation}
	where $\alpha$ is a dimensionless constant whose value is not confirmed yet and
  the parameter $K_{\alpha}$ that depends on
the power $\alpha$ of the entropy correction as following:
	\begin{equation*}
	K_{\alpha} = \frac{\alpha (4\pi)^{\frac{\alpha}{2}-1}}{(4 - \alpha)r_c^{2 - \alpha}},
	\end{equation*}
	here $r_c$ denotes crossover scale. 	The  Boltzmann constant is set unit here. 
		 The last term of equation (\ref{en}) appears due to the superposition of ground state and excited state  wave function of the field.  
Interestingly,	Verlinde \cite{5} proposed an entropic origin of gravity and interpreted gravity as an entropic
force. The gravity derived here was from the notion of information associated with matter and its location measured in terms of entropy.
		 The  Newton's law of  gravitational force $F$ is related to the entropy $S$ of the system as \cite{12}
		\begin{eqnarray}
		 F = -4 l_p^2 \frac{GM^2}{r^2} \frac{\partial S}{\partial A}.
\end{eqnarray}		
For the power-law corrected  entropy given in  (\ref{en}), this relation leads to  the following modified Newton's force \cite{12}: 
	\begin{equation}
F = -\frac{GM^2}{r^2}\left[1-\frac{\alpha}{2}\left(\frac{r_c}{r}\right)^{\alpha - 2}\right],
\label{nn}
	\end{equation} 
	which coincides with the original 
 Newtons law when $\alpha$ is set to zero. Keeping in view the attractive nature of gravity, we should have $F<0$. This requires,
\begin{equation*}
\alpha<2\left(\frac{r}{r_c}\right)^{\alpha-2}.
\end{equation*}

	In this paper, we study the effect of power-law corrected Newtonian potential on the clustering of galaxies. We also consider the effect of dark matter on the formation of galaxy clusters via the incorporation of cosmological constant in the Newtonian potential. This is because of the role played by cosmological constant $\Lambda$  in the expansion of universe  ~\cite{5}. In order to study the effects of all the corrections made in gravitational potential in galaxy clustering, we first derive the $N$-body partition function by evaluating configuration integrals recursively. The resulting partition function is employed to calculate the various thermodynamic quantities viz the Helmholtz free energy, entropy, pressure, internal energy, and chemical potential which possess deviations from their original values due to the incorporation of corrections. Remarkably, a modified clustering parameter emerges naturally from the corrected equations of state. Furthermore, we derive the probability distribution function assuming that the system is in a quasi equilibrium state as described by the grand canonical ensemble. The  expression of distribution function embeds modified clustering parameter. A comparative analysis is made with the original distribution function to study the deviation due to corrections. 
	
	The paper is organized as following. In section 2, we  consider a power-law and cosmological 
	constant modified gravitational potential to calculated partition function of the system of galaxies and galaxy clusters. With the help of resulting partition function, the various thermodynamical equation of states are calculated in section 3. The study the distribution of galaxies and galaxy's clusters under the modified gravitational potential, we estimate distribution function in section 4. Finally, we draw concluding remarks in the last section. 
	\section{Interaction of galaxy clusters under modified potential}
 	In this section, we derive a modified Newtonian potential due to power-law corrected force   and   estimate the partition function.  
	\subsection{A modified gravitation potential}
 Utilized standard definition   $\Phi=-\int Fdr$,  we calculate a power-law  modified gravitational potential corresponding to power-law corrected Newton's law (\ref{nn})  as
\begin{equation}
		\Phi=-GM^2\left[\frac{1}{r}+\frac{\alpha}{2\left(\alpha +1\right)}r_c^{\alpha-2}r^{\alpha+1}\right].
\end{equation}
This potential  can arise in modified gravity theories like $f(R)$ gravity.
In fact, the power-law entropy corrected Friedmann equation is 
derived using the first-law on the apparent horizon \cite{12}.

 To get more realistic results, one can not ignore the cosmological constant term 
$−\frac{1}{6}\Lambda r$ \cite{13}
in potential at the cosmological scale  as
cosmological constant  is responsible for the expansion
of the Universe through a repulsive force. 
Therefore,  the exact gravitation potential is given by
\begin{equation}
		\Phi=-GM^2\left[\frac{1}{r}+\frac{\alpha}{2\left(\alpha +1\right)}r_c^{\alpha-2}r^{\alpha+1}+\frac{1}{6}\frac{\Lambda r^2}{GM^2}\right].
\end{equation} 
This is a final gravitational potential where second and third terms correspond to the power-law and cosmological constant correction terms respectively.

\subsection{Generating functional of galaxies cluster under modified gravity}

Next, we estimate  the partition function for the system of galaxies under 
the modified gravitational potential. Here it is assumed that  system of galaxies follows 
a statistically homogeneous distribution
over large regions, which consists of an ensemble
of cells with equal volume $V$ and   equal average
density.  Let us begin by writing  the general partition function for the system comprised with $N$ galaxies of equal  mass $M$, momenta $P_i$ and average temperature $T$  as
\begin{equation}
Z_N\left(T,V\right)=\dfrac{1}{\lambda ^{3N} N!} \int d^{3N}P d^{3N}r\ \exp\left[  -\frac{1}{T}\left(\sum_{i=1}^{N}\frac{P_i^2}{2M}+\Phi\left(r_1,r_2,...r_N\right)\right)\right] ,
\end{equation} 
where $N!$ appears due to  distinguishable  nature of galaxies and   $\lambda$ is a normalization constant  for the phase space volume cell.
Upon integration over momentum space, this further simplifies to
\begin{equation}
	Z_N\left(T,V\right)=\frac{1}{N!}\left(\frac{2\pi MT}{\lambda ^2}\right)^{3N/2}Q_N\left(T,V\right),\label{zn}
\end{equation}
where the configuration integral, $Q_N\left(T,V\right)$, has the following form: 
\begin{equation}
	Q_N\left(T,V\right)=\int...\int\prod_{1\leq i<j\leq N}\left(1+\frac{\Phi}{T}\right)d^{3N} r,
\end{equation} 
 Here we   neglected the higher-order terms of potential  as the system of galaxies is still clustering.

Here we note that for the point-mass  galaxies (i.e., for $r= 0$), 
the potential energy and consequently partition
function diverges. In order to remove this discrepancy, we assume galaxies of 
  extended nature (i.e., galaxies with halos). For this we introduce a softening parameter  $\epsilon$, which assures  the finite size of galaxies. The value of this softening parameter ranges $0.01\leq \epsilon \leq 0.05$.
 In order to estimate  partition function, the modified potential incorporates softening parameter appropriately as follows,
	\begin{equation}
		\Phi \left(\epsilon\right)=-GM^2\left[\frac{1}{\left(r^2+\epsilon ^2\right)^{\frac{1}{2}}}+\frac{\alpha r_c^{\alpha-2}}{2\left(1-\alpha\right)}r^{1-\alpha}+\frac{1}{6}\frac{\Lambda r^2}{GM^2}\right].\label{ph}
	\end{equation}
	The second and third terms do not require softening parameter to be introduces as corresponding potentials do not diverge.
 
	 Now, we estimate configuration integral $Q_N\left(T,V\right)$ iteratively.
	The configuration integral for a single (spherically) galaxy  of radius $R$ is given by 
	\begin{equation}
		Q_1\left(T,V\right)=V.
	\end{equation}
	For system of two galaxies, the configuration integral is given by
	\begin{equation}
		Q_2\left(T,V\right)=4\pi V\int dR\  R^2\left[1+\frac{GM^2}{T}\left(\frac{1}{\left(R^2+\epsilon ^2\right)^{\frac{1}{2}}}+\frac{\alpha r_c^{\alpha-2}}{2\left(1-\alpha\right)}R^{1-\alpha}+\frac{1}{6}\frac{\Lambda R^2}{GM^2}\right)\right].\label{q}
	\end{equation}
	Here,   the double integral reduces to a single
integral by fixing  the position of one  galaxy.
	This result can also be obtained by considering the fact that expansion of universe exactly cancels the effect of the long-range mean gravitational field on the particle motions \cite{15}.
Eq. (\ref{q}) further simplifies to
	\begin{eqnarray}
		Q_2\left(T,V\right)&=&V^2\left[1+\frac{3}{2}\frac{GM^2}{RT}\left(\sqrt{1+\frac{\epsilon^2}{R^2}}+\frac{\epsilon^2}{R^2}\ln \frac{ {\epsilon}/{R}}{1+\sqrt{1+\frac{\epsilon^2}{R^2}}} \right.\right.\nonumber\\
		&+&\left.\left. \frac{3\alpha}{2\left(1-\alpha\right)\left(4-\alpha\right)}r_c^{\alpha-2}R^{1-\alpha}+\frac{\Lambda R^3}{15GM^2}\right) \right].
	\end{eqnarray}  
	More compactly, this can be written   as 
	\begin{equation}
		Q_2\left(T,V\right)=V^2\left[1+\frac{3}{2}\left(\zeta+\gamma+\beta\right)\frac{GM^2}{RT}\right],\label{q2}
	\end{equation}
	where
	\begin{align*}\zeta&=\sqrt{1+\frac{\epsilon^2}{R^2}}+\frac{\epsilon^2}{R^2}\ln \frac{ {\epsilon}/{R}}{1+\sqrt{1+\frac{\epsilon^2}{R^2}}} ,\\
	\gamma&=\frac{3\alpha}{2\left(1-\alpha\right)\left(4-\alpha\right)}r_c^{\alpha-2}R^{1-\alpha},\\
	\beta &=\frac{\Lambda R^3}{15GM^2}.
	\end{align*} 
	\par
	Here, we note that $\alpha$ in the $\gamma$ takes the value  $0.624 < \alpha < 2$ because within this limit only   a phantom accelerating universe can be derived which is compatible with the observations \cite{k}.

	Next, we scale the temperature $T$ and radius $R$ as
	 $T\rightarrow\eta^{-1}T $ and	 $R\rightarrow\eta R$, which 
 leads to the dimensionless factor $\frac{GM^2}{RT}$ scale invariant. Therefore, we can scale   $\frac{GM^2}{RT}\rightarrow\left(\frac{GM^2}{RT}\right)^3=\frac{3}{2}\left(\frac{GM^2}{T}\right)^3\bar{\rho} := x$ \cite{15a}. 
Thus, Eq. (\ref{q2}) finally reduces to
	\begin{equation}
			Q_2\left(T,V\right)=V^2\left[1+\left(\zeta+\gamma+\beta\right)x\right].
	\end{equation}	
Following this procedure iteratively, the configuration integral for $N$ galaxies is obtained as
\begin{equation}
	Q_N\left(T,V\right)=V^N\left[1+\left(\zeta+\gamma+\beta\right)x\right]^{N-1}.
\end{equation}
 By inserting this value of  configuration integral of $N$ galaxies into the the partition function  (\ref{zn}), we achieve the expression of partition function  for  $N$ galaxies 
 gravitating under modified gravity as
\begin{equation}
	Z_N\left(T,V\right)=\frac{1}{N!}\left(\frac{2\pi MT}{\lambda ^2}\right)V^N\left[1+\left(\zeta+\gamma+\beta\right)x\right]^{N-1}.\label{part}
\end{equation}

\section{Thermodynamical equations of state} 
In this section, we derive various more exact equations of states for the system of galaxies 
interacting under modified potential.  More precisely, we  derive the Helmholtz free energy, entropy, internal energy, pressure  and chemical potential. We also emphasize the effect of corrections
in Newton's potential on these equation of states. 
\subsection{Helmholtz free energy }
The Helmholtz free energy can be calculated from the partition function using following definition:
\begin{equation}
	F=-T\ln Z_N(T,V).
\end{equation}
Therefore, the Helmholtz free energy for our system of galaxies is calculated by
 \begin{eqnarray}
	F=NT\left(\ln\frac{N}{V}T^{-3/2}\right)-NT-NT\ln\left[1+\left(\zeta+\gamma+\beta\right)x\right]-\frac{3}{2}NT\ln\left(\frac{2\pi M}{\lambda^2}\right).
  \end{eqnarray}
 Here we have made following assumption   $(N-1)\approx N$, since  $N$ is very large.
  This equation can also be written as
  \begin{equation}
  F=NT\left(\ln\frac{N}{V}T^{-3/2}\right)-NT+NT\ln\left[1-\frac{(\zeta+\gamma+\beta)x}{1+(\zeta+\gamma+\beta)x}\right]-
  \frac{3}{2}NT\ln\left(\frac{2\pi M}{\lambda^2}\right).\label{hel}
  \end{equation}
   \begin{figure}[htb]
 $%
\begin{array}{cc}
\epsfxsize=7cm \epsffile{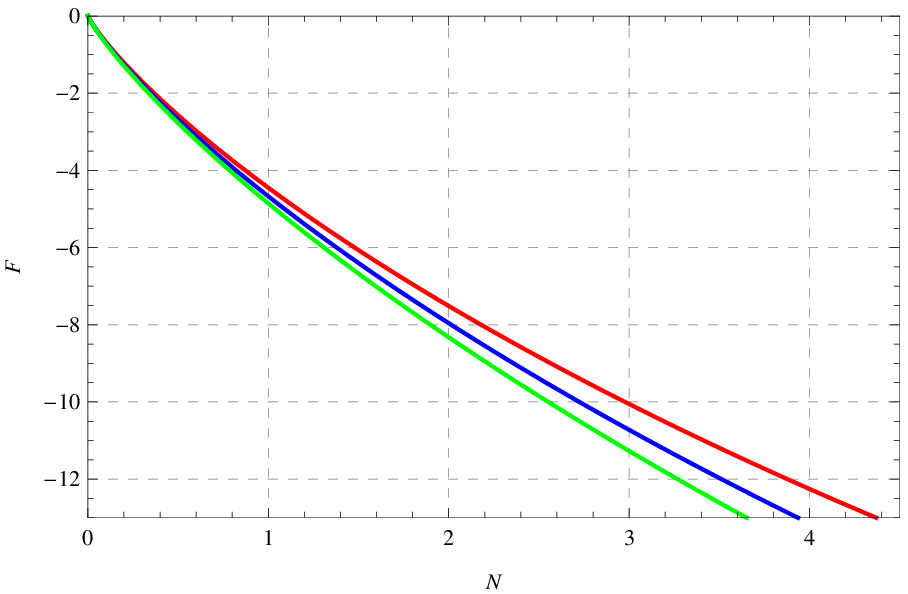} &\ \ \ \ \ \ \epsfxsize=7cm %
\epsffile{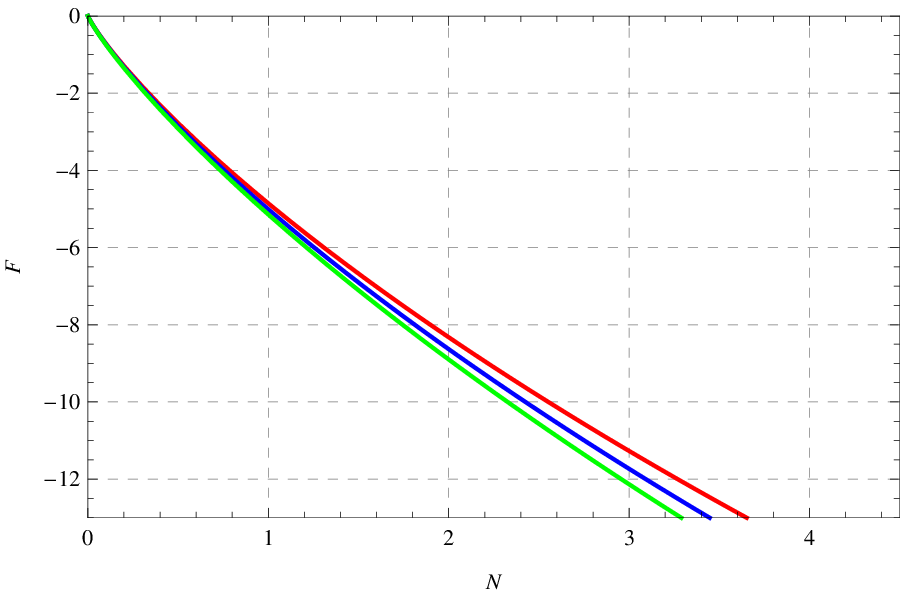} %
\end{array}
$%
 \caption{Helmholtz free energy vs. number of galaxies.
	 Left:  red, blue, and green lines correspond to $\gamma x  = 0, 0.5$ and $ 1$, respectively,  with $\zeta x  =1$, $\beta x  =0$. Right:    red, blue, and green lines correspond to $\gamma x  = 0, 0.5$ and $ 1$, respectively, with  $\zeta x  =1$, $\beta x  =1$.  Rest of the parameters  are unit here.
 }\label{fig1}
 \end{figure} 
     \begin{figure}[htb]
 $%
\begin{array}{c}
\epsfxsize=7cm \epsffile{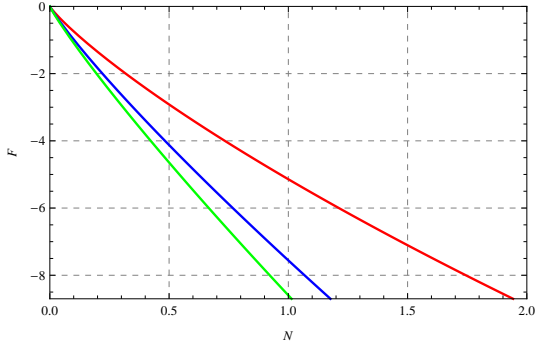}  %
\end{array}
$%
 \caption{Behavior of Helmholtz free energy versus particle number  for different values of $M$. We  set  all the parameters along with $\zeta x$,  $\beta x$  and $\gamma x$ to unit. Here, red, blue, and green lines correspond to  $M=1$, $M=5$  and  $M=10$, respectively.
 }\label{fig2}
 \end{figure}
 In order to study the behavior of Helmholtz free energy with respect to particle number, we plot a graph \ref{fig1} and  \ref{fig2}. From the figure, we see that Helmholtz free energy is a decreasing 
 function of $N$. It means as long as number of galaxies increases the Helmholtz free energy decreases. It is also evident that the corrections due to power-law and cosmological constant terms make the   free energy more negative. Also, the free energy decreases with increase in 
 the mass of galaxies.
 
 By comparing this equation of state  to its original form given in Ref. \cite{15}, we 
conclude that the corrections in Newton's potential are apparent in clustering parameter only.
The new clustering parameter corresponding to power-law  and cosmological constant corrections
is obtained as 
  \begin{equation}
b_\star =\frac{(\zeta+\gamma+\beta)x}{1+(\zeta+\gamma+\beta)x}.
  \end{equation}
  This  modified clustering parameter can be expressed  in terms of   original clustering parameter
   $b_\epsilon=\frac{\zeta x}{1+\zeta x}$ \cite{15} as follows
  \begin{equation}
b_\star=\frac{b_\epsilon(1-\gamma x-\beta x)+(\gamma+\beta)x}{1+(\gamma+\beta)x-b_\epsilon (\gamma+\beta)x}.
  \end{equation} 
 
    \subsection{Entropy}
Let us calculate entropy of the system which a very important thermodynamical quantity.
For a given Helmholtz free energy (\ref{hel}), the entropy can easily be calculated as
 \begin{eqnarray}
	S=-N\ln\left(\frac{N}{V}T^{-3/2}\right)-N\ln[1-b_\star]-3Nb_\star+
	\frac{5}{2}N+\frac{3}{2}N\ln \frac{2\pi M}{\lambda^2}.\label{ent}
\end{eqnarray}
Here we have utilized standard entropy definition $	S=-\left(\frac{\partial F}{\partial T}\right)_{N,V}$.
The specific entropy (entropy per galaxy) is given by
\begin{equation}
	\frac{S}{N}=-\ln\left(\frac{N}{V}T^{-3/2}\right)-\ln[1-b_\star]-3b_\star +S_0,
\end{equation}
where $S_0 =\frac{5}{2}+\frac{3}{2}\ln \frac{2\pi M}{\lambda^2}$ is a constant.
 \begin{figure}[htb]
 $%
\begin{array}{cc}
\epsfxsize=7cm \epsffile{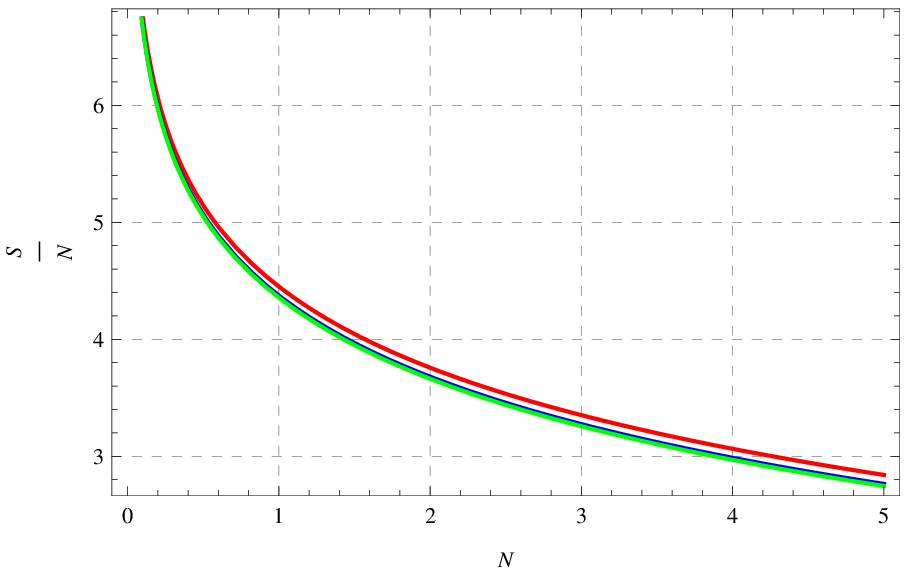} &\ \ \ \ \ \ \epsfxsize=7cm %
\epsffile{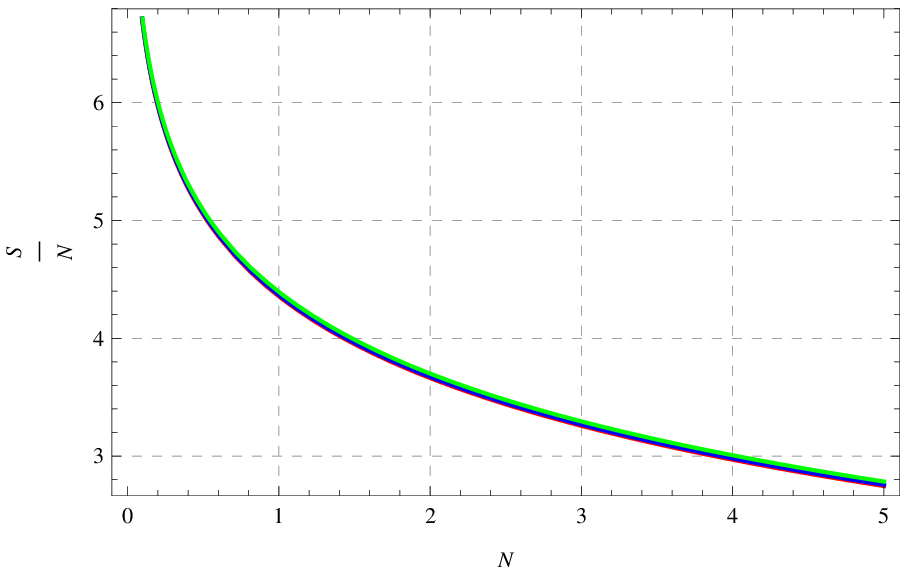} %
\end{array}
$%
 \caption{Specific entropy ($S/N$) versus particle number  ($N$)  with (right) and without (left) dark energy contributions. Left:  red, blue, and green lines correspond to $\gamma x  = 0, 0.5$ and $ 1$, respectively,  with $\zeta x  =1$ and $\beta x  =0$. Right:    red, blue, and green lines correspond to $\gamma x  = 0, 0.5$ and $ 1$, respectively with  $\zeta x  =1$ and $\beta x  =1$.  Rest of the parameters  are unit here.}\label{fig3}
 \end{figure}
     \begin{figure}[htb]
 $%
\begin{array}{c}
\epsfxsize=7cm \epsffile{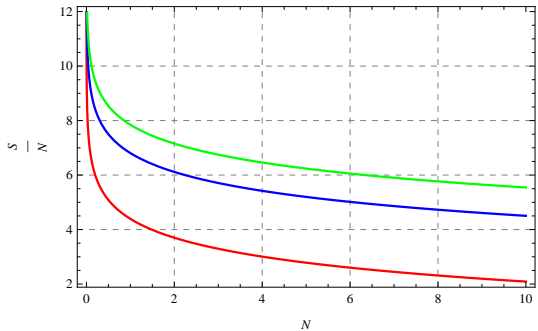}  %
\end{array}
$%
 \caption{Behavior of specific entropy ($S/N$) versus particle number  ($N$) for different values of $M$. We  set  all the parameters along with $\zeta x$,  $\beta x$  and $\gamma x$ to unit. Here, red, blue, and green lines correspond to  $M=1$, $M=5$  and  $M=10$, respectively. }\label{fig4}
 \end{figure}
 The behavior of  specific entropy with respect to number of galaxies can be seen in Figs. \ref{fig3} and \ref{fig4}. The specific entropy is also a positive valued decreasing function  with number of galaxies. The power-law corrected term makes specific entropy smaller in absence of dark energy. However, the effect of power-law correction is not significant in the presence of dark energy but makes specific entropy bit positive. The specific entropy 
increases significantly along with increase in mass of galaxies.
 \subsection{Internal Energy}
The internal energy of the system is defined as $U=F+TS$. For a given Helmholtz free energy (\ref{hel}) and entropy (\ref{ent}), the internal energy is calculated by
\begin{equation}
	U =\frac{3}{2}NT\left[1-2\frac{\left(\zeta+\gamma+\beta\right)x}{1+\left(\zeta+\gamma+\beta\right)x}\right] 
	 =\frac{3}{2}NT\left[1-2b_\star \right].
\end{equation}
\begin{figure}[htb]
 $%
\begin{array}{cc}
\epsfxsize=7cm \epsffile{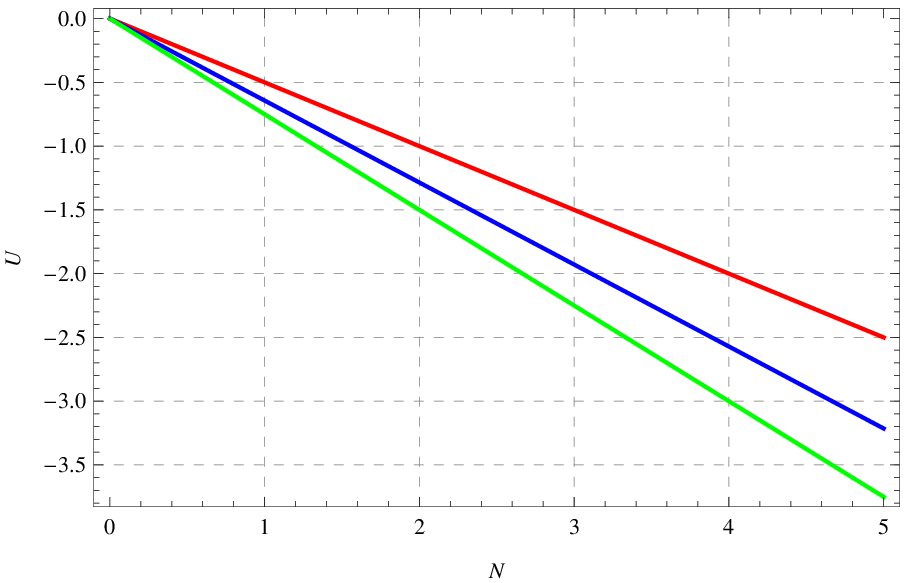} &\ \ \ \ \ \ \epsfxsize=7cm %
\epsffile{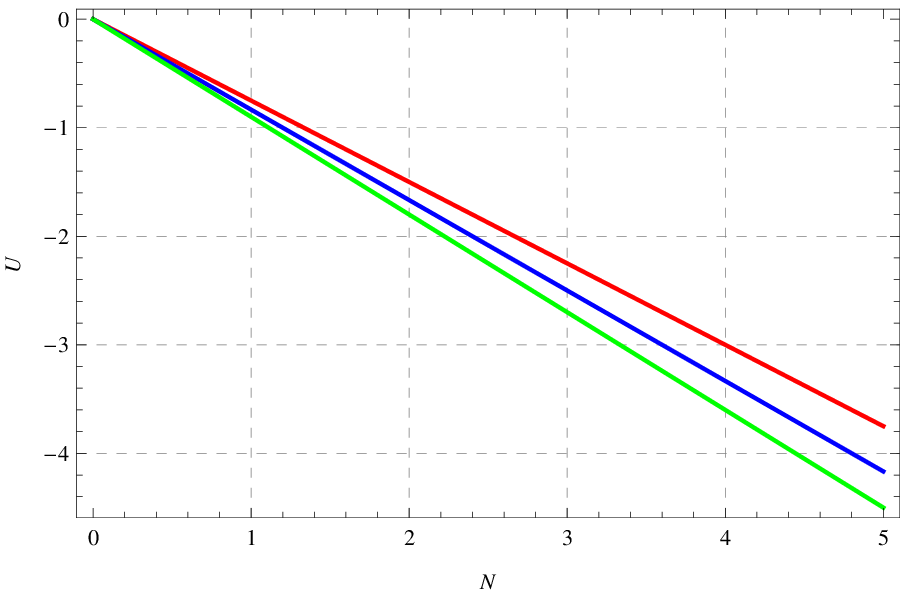} %
\end{array}
$%
 \caption{Internal energy ($U$) versus particle number  ($N$)  with (right) and without (left) dark energy contributions. Left:  red, blue, and green lines correspond to $\gamma x  = 0, 0.5$ and $ 1$, respectively,  with $\zeta x  =2$ and $\beta x  =0$. Right:    red, blue, and green lines correspond to $\gamma x  = 0, 0.5$ and $ 1$, respectively with  $\zeta x  =2$ and $\beta x  =1$.  Rest of the parameters  are unit here.}\label{fig5}
 \end{figure}
 From the figure \ref{fig5}, we observe that the internal energy of the system which is negative valued becomes more negative under the effect of correction in gravitational potential. 
\subsection{Pressure}
The standard definition of pressure is given by  $P=-\left(\frac{\partial F}{\partial V}\right)_{N,T}.$ Exploiting expression (\ref{hel}), the pressure for a gravitating system under
modified gravity is given by
\begin{equation}
	P=\frac{NT}{V}\left[1-b_\star\right]. 
  \end{equation}
  \begin{figure}[htb]
 $%
\begin{array}{cc}
\epsfxsize=7cm \epsffile{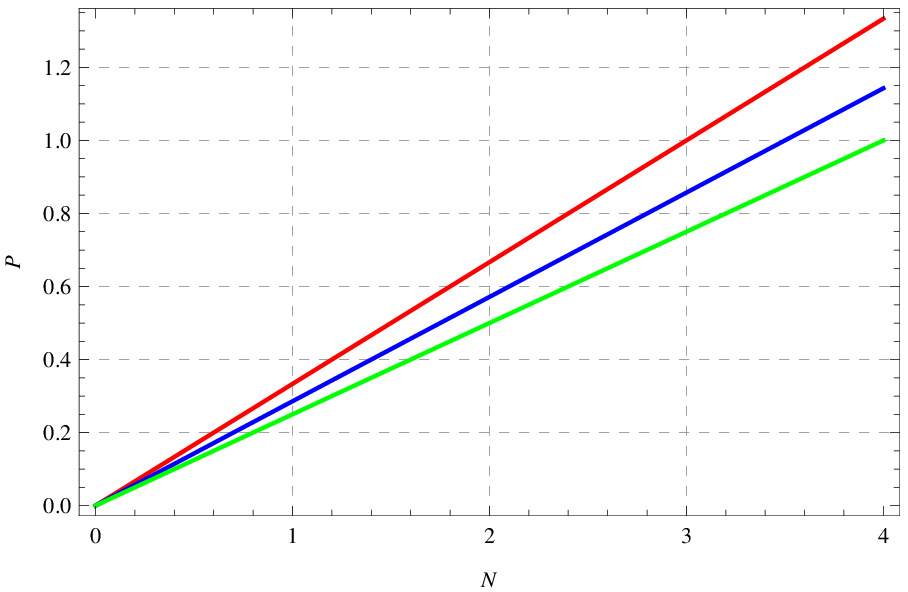} &\ \ \ \ \ \ \epsfxsize=7cm %
\epsffile{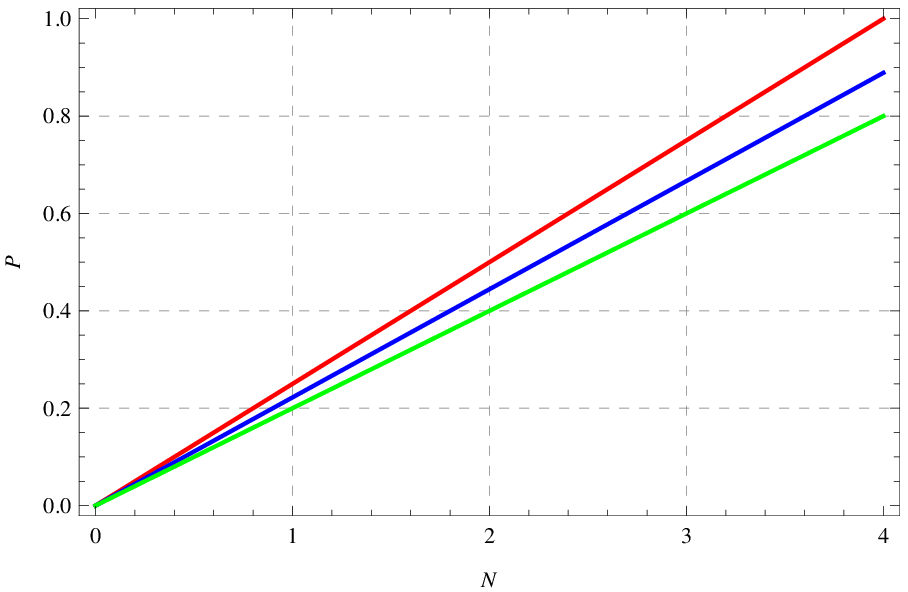} %
\end{array}
$%
 \caption{Pressure ($P$) versus particle number  ($N$)  with (right) and without (left) dark energy contributions. Left:  red, blue, and green lines correspond to $\gamma x  = 0, 0.5$ and $ 1$, respectively,  with $\zeta x  =2$ and $\beta x  =0$. Right:    red, blue, and green lines correspond to $\gamma x  = 0, 0.5$ and $ 1$, respectively with  $\zeta x  =2$ and $\beta x  =1$.  Rest of the parameters  are unit here.}\label{fig6}
 \end{figure}
 The pressure is a linear function of number of the galaxies, which means that pressure increases as long as number of galaxies increases. It is obvious from the figure \ref{fig6} that the presence of correction terms negate the 
 pressure of the system. 
\subsection{Chemical Potential}
The chemical potential   which measure exchange of galaxies can be calculated from the formula $\mu=-\left(\frac{\partial F}{\partial N}\right)_{V,T} $. So, it is a matter of calculation 
to derive chemical potential for this system as
\begin{equation}
	\mu = T\left(\ln \frac{N}{V}T^{-3/2}\right)-T\ln \left[1+(\zeta+\gamma+\beta)x\right]-T\frac{(\zeta+\gamma+\beta)x}{1+(\zeta+\gamma+\beta)x}
-\frac{3}{2}T\ln\left(\frac{2\pi M}{\lambda^2}\right).
\end{equation}
This further simplifies to
\begin{equation}
\mu=T\left(\ln \frac{N}{V}T^{-3/2}\right)+T\ln \left[1-b_\star\right]-Tb_\star
-\frac{3}{2}T\ln\left(\frac{2\pi M}{\lambda^2}\right).\label{che}
 \end{equation}
  \begin{figure}[htb]
 $%
\begin{array}{cc}
\epsfxsize=7cm \epsffile{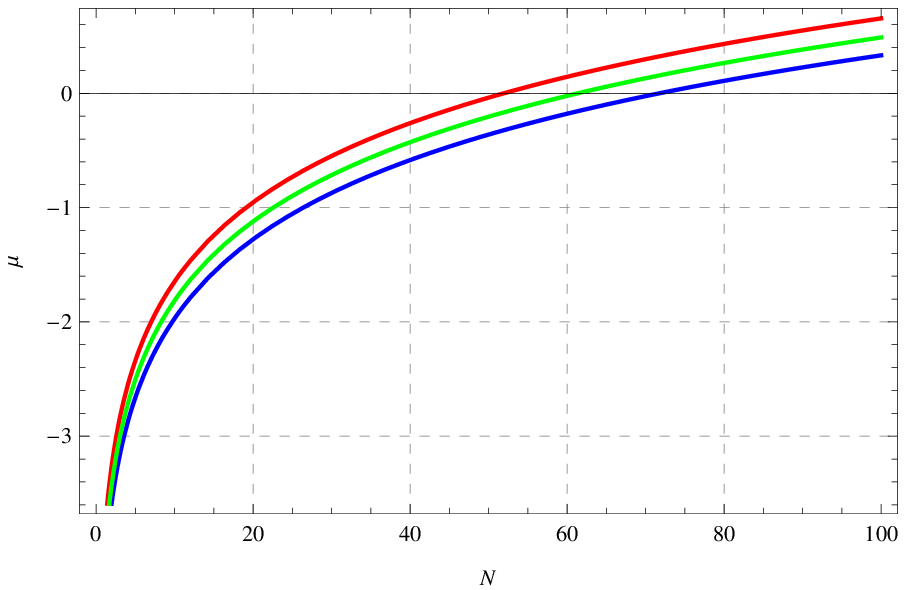} &\ \ \ \ \ \ \epsfxsize=7cm %
\epsffile{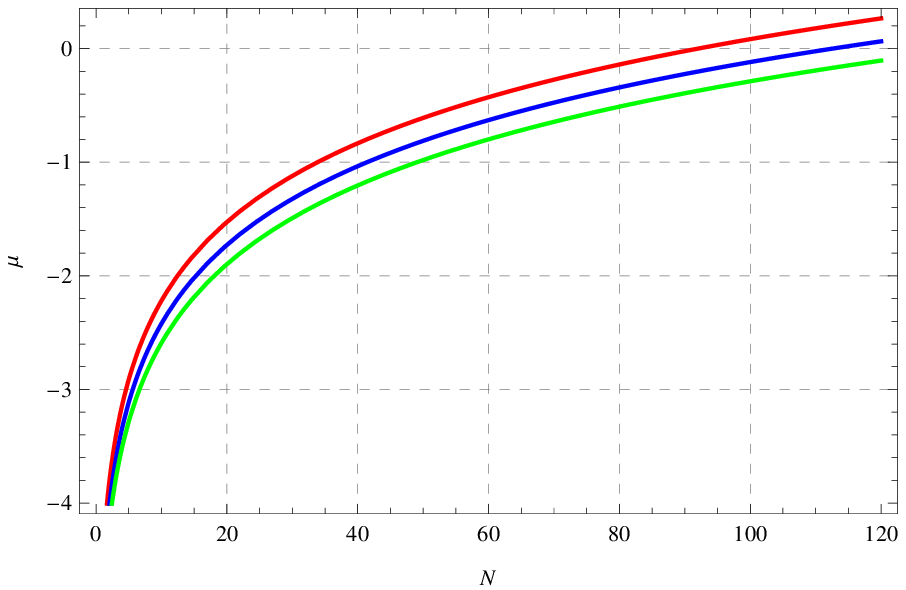} %
\end{array}
$%
 \caption{Chemical potential ($\mu(N)$) versus particle number  ($N$)  with (right) and without (left) dark energy contributions. Left:  red, blue, and green lines correspond to $\gamma x  = 0, 0.5$ and $ 1$, respectively,  with $\zeta x  =1$ and $\beta x  =0$. Right:    red, blue, and green lines correspond to $\gamma x  = 0, 0.5$ and $ 1$, respectively with  $\zeta x  =1$ and $\beta x  =1$.  Rest of the parameters  are unit here.}\label{fig7}
 \end{figure}
     \begin{figure}[htb]
 $%
\begin{array}{c}
\epsfxsize=7cm \epsffile{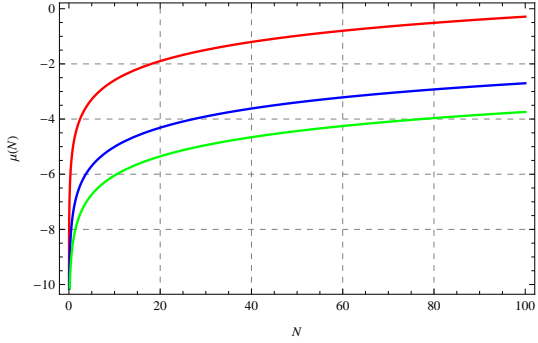}  %
\end{array}
$%
 \caption{Behavior of chemical potential ($\mu(N)$) versus particle number  ($N$) for different values of $M$. We  set  all the parameters along with $\zeta x$,  $\beta x$  and $\gamma x$ to unit. Here, red, blue, and green lines correspond to  $M=1$, $M=5$  and  $M=10$, respectively. }\label{fig8}
 \end{figure}
The behavior of chemical potential with respect to number of galaxies can be seen from figures \ref{fig7} and \ref{fig8}. Interestingly, we observe that the chemical potential is negative 
valued for the system of galaxies with small number of galaxies. However, it becomes positive valued when  number of galaxies increases to a specific value. The presence of correction 
terms in clustering parameter decreases the chemical potential of the system. The chemical potential decreases with increase of the mass of galaxies.

At the end of this section, we conclude that  the modified Newton's potential amounts changes to the  clustering parameter $b_\star$ only, however   the basic structures of the equations
remain intact.  
\section{General distribution function}
In order to find the probability distribution function $F(N)$, which contains void distribution as well as statistics of counts  of the number of  galaxies  in cells throughout the system.  For the system of galaxies, wherein galaxies as well as energy can cross the cell boundary, one has to estimate the grand canonical partition function. The grand canonical partition function,   a weighted sum of all canonical partition functions, is defined by
 \begin{equation}
Z_G(T,V,z)=\sum_{N=0}^{\infty} e^{\frac{N\mu}{T}}Z_N(T,V),
\end{equation}  
where $z$ is the fugacity. The grand partition function for the system of galaxies is expressed  in terms of thermodynamic variables  as
\begin{equation}
\ln Z_G=\frac{PV}{T}=N(1-b_\star).\label{zg}
\end{equation} 
 The probability distribution function $F(N)$ for finding $N$ galaxies in a cell of volume $V$ and energy $U(N,V)$ is defined by
\begin{equation}
F(N)=\frac{\sum_i e^{\frac{N\mu}{T}}e^{\frac{U_i}{T}}}{Z_G(T,V,z)}=\frac{e^{\frac{N\mu}{T}}Z_N(V,T)}{Z_G(T,V,z)}.
\end{equation}
Making use of Eqs. (\ref{part}),  (\ref{che})  and (\ref{zg}), the distribution function for the extended mass galaxies under modified potential is  estimated as
\begin{equation}
F(N)=\frac{\bar{N}}{N!}(1-b_\star)[\bar{N}(1-b_\star)+Nb_\star]^{N-1}e^{-Nb_\star-\bar{N}(1-b_\star)}.
\end{equation} 
This distribution function is structurally similar to those derived originally  by Saslaw and Hamilton  ~\cite{16} from thermodynamic point of view and by Ahmad and Saslaw ~\cite{15} from statistical point of view. Also the distribution function derived from  logarithmic and volume corrected Newtonian potential ~\cite{10} has the same general structure.
 \begin{figure}[htb]
 $%
\begin{array}{cc}
\epsfxsize=7cm \epsffile{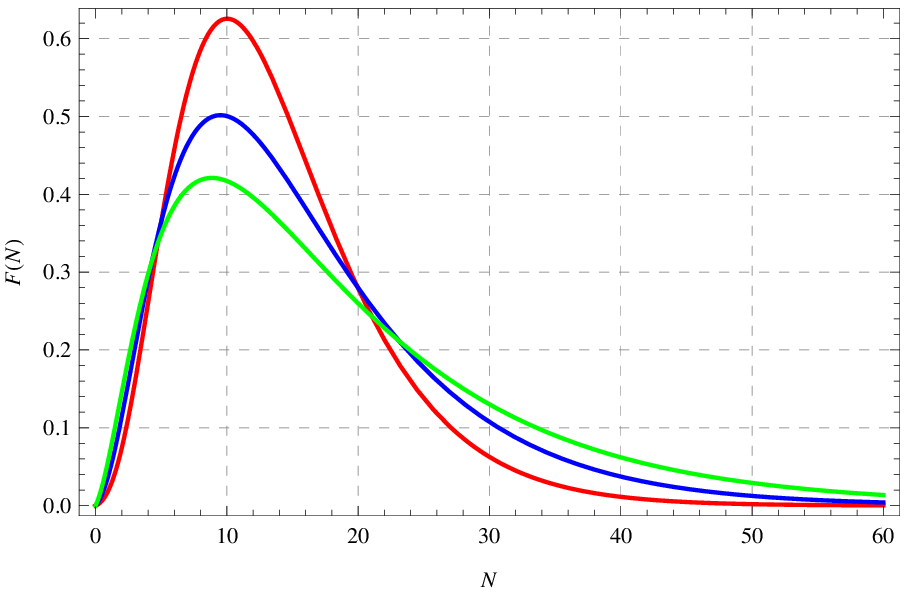} &\ \ \ \ \ \ \epsfxsize=7cm %
\epsffile{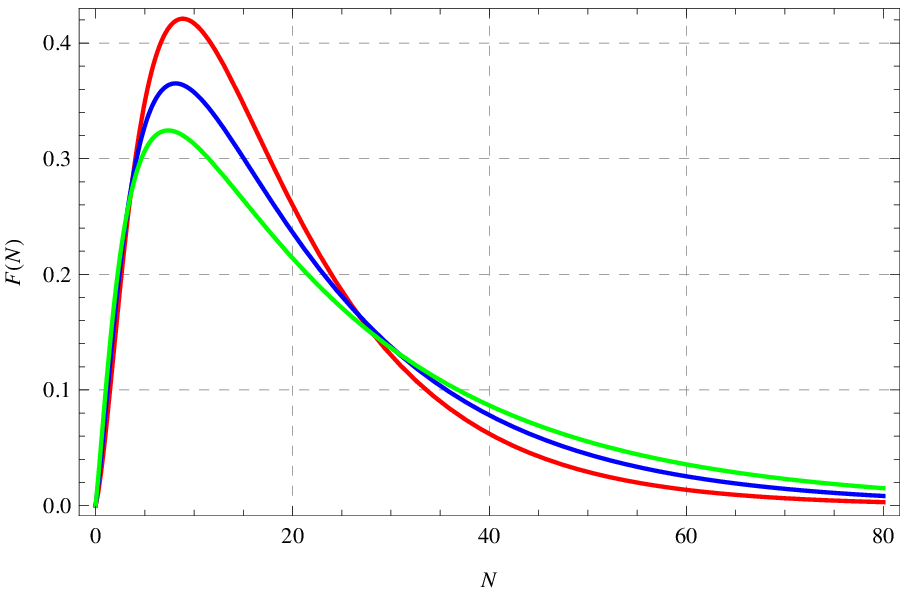} %
\end{array}
$%
 \caption{The distribution function $F(N)$ versus $N$ for $\bar N=10$ with (right) and without 
  (left) cosmological constant  contributions. Left:  red, blue, and green lines correspond to 
  $\gamma x  = 0, 0.5$ and $ 1$, respectively,  with $\zeta x  =1$ and $\beta x  =0$. Right:    red, blue, and green lines correspond to $\gamma x  = 0, 0.5$ and $ 1$, respectively with  $\zeta x  =1$ and $\beta x  =1$.}\label{fig9}
 \end{figure}
 The behavior of distribution function $F(N)$ versus $N$ can be seen from the comparative analysis   as given in Fig. \ref{fig9}.  The presence of correction term decreases the 
peak value of  distribution function which occurs for system of small number of galaxies.
 However, for the system of large number of galaxies  the correction terms increase the distribution function. 
 
\section{Discussion and conclusions}
We have presented a study of galaxy clustering under the modified Newton's law. These modifications to Newton's law incorporate  a power-law entropic corrections  along with the inclusion of cosmological constant $\Lambda$ term, in order to take into account the effect of quantum entanglement (a possible source of black hole entropy) and   dark energy respectively. Utilizing the modified Newtonian potential, we have derived the corresponding canonical partition function for the system of extended mass galaxies with the assumption that the system is made of $N$ equal volume cells and average particle density $\bar{\rho}$, which are statistically homogeneous over large regions.  In this regard, we  have used of softening parameter $\epsilon$ in the first term of Hamiltonian to get rid of the divergence of the Hamiltonian when galaxies are considered as point like. This justifies that  the extended nature (finite size) of  galaxies (or galaxies with halos). From the resulting partition function, we have calculated various important thermodynamic equations of state. Namely, these are free energy, entropy, internal energy, pressure and chemical potential. The  exact expressions of equations of state contain  a corrected correlation (clustering) parameter which emerges naturally for the clusters of the galaxies with halos. The new clustering parameter $b_\star$ reduces to the original parameter $b_\epsilon$ when $\gamma =0$ and $\beta=0$. Moreover,  the  distribution function is modified due to correction in potential but has the similar structure as the original one.      

  \end{document}